\documentclass[%
 aip,
 jmp,%
 amsmath,amssymb,
 reprint,%
]{revtex4-1}

\usepackage{graphicx}
\usepackage{dcolumn}
\usepackage{bm}
\usepackage{multirow}

\begin{document}

\preprint{AIP/123-QED}

\title[Real-time feedback from iterative electronic structure calculations]{Real-time feedback from iterative electronic structure calculations}

\author{Alain C. Vaucher}
\affiliation{ 
ETH Z\"urich, Laboratorium f\"ur Physikalische Chemie, Vladimir-Prelog-Weg 2, CH-8093 Z\"urich, Switzerland
}%
\author{Moritz P. Haag}%
\affiliation{ 
ETH Z\"urich, Laboratorium f\"ur Physikalische Chemie, Vladimir-Prelog-Weg 2, CH-8093 Z\"urich, Switzerland
}%

\author{Markus Reiher}
 \email{markus.reiher@phys.chem.ethz.ch}
\affiliation{ 
ETH Z\"urich, Laboratorium f\"ur Physikalische Chemie, Vladimir-Prelog-Weg 2, CH-8093 Z\"urich, Switzerland
}%

\date{\today}

\begin{abstract}
Real-time feedback from iterative electronic structure calculations requires to mediate between the inherently unpredictable execution times of the iterative algorithm employed and the necessity to provide data in fixed and short time intervals for real-time rendering. 
We introduce the concept of a mediator as a component able to deal with infrequent and unpredictable reference data to generate reliable feedback.
In the context of real-time quantum chemistry, the mediator takes the form of a surrogate potential that has the same local shape as the first-principles potential and can be evaluated efficiently to deliver atomic forces as real-time feedback.
The surrogate potential is updated continuously by electronic structure calculations and guarantees to provide a reliable response to the operator for any molecular structure. 
To demonstrate the application of iterative electronic structure methods in real-time reactivity exploration, we implement 
self-consistent semi-empirical methods as the data source and apply the surrogate-potential mediator to deliver reliable real-time feedback.

\end{abstract}

\maketitle

\setlength{\parindent}{0cm}
\setlength{\parskip}{0.6em plus0.2em minus0.1em}

\section{Introduction}

Knowledge of the atomic rearrangements occurring during chemical reactions is crucial for understanding reactivity. 
It allows designing new reactions and optimizing existing ones.
In recent years, approaches toward interactive reactivity studies\cite{haag2014a} have been introduced. 

We developed Haptic Quantum Chemistry\cite{marti2009,haag2011} and Real-Time Quantum Chemistry\cite{haag2013} to actively interact with 
some molecular system under consideration and perceive its immediate (quantum-mechanical) response. 
This immersion into the molecular world is enabled by immediate feedback of quantum-chemical calculations, which is of paramount importance to guide operators during the (interactive) exploration of molecular systems.
Interactivity enables chemists to understand reactivity more intuitively and more efficiently than with traditional tools.
In other approaches toward the interaction with molecular systems, the emphasis is on the possibility to steer simulations in real time.
For example, in the field of interactive molecular dynamics, Stone et al.\ allowed operators to drive classical molecular dynamics simulations toward events that would happen too rarely or not at all otherwise\cite{stone2001}. 
This field was recently extended to \textit{ab initio} molecular dynamics by Luehr et al.\cite{luehr2015a}
Another approach was implemented by Bosson et al.,\cite{bosson2012} who applied a non-iterative
semi-empirical quantum-chemical method that instantaneously optimizes a molecular structure set up with their molecular
editor {\sc Samson}\cite{samson}. Such structure relaxation approaches to quickly generate reasonable molecular structures (often based on classical
force fields) are becoming a standard tool for structure generation attempts in such graphical user interfaces (cf.\ for another example the
{\sc Avogadro} program \cite{hanwell2012}).

Real-time quantum chemistry\cite{haag2013} allows exploring the Born--Oppenheimer potential energy surface interactively in real time to gain intuitive insight into reactivity.
Such real-time reactivity explorations rely on two types of feedback.
Firstly, operators can actively manipulate the molecular structure, for example, with a haptic device that allows them
to move atoms while experiencing the quantum-chemical force acting on them. 
The haptic feedback provides an immediate and intuitive understanding of which parts of the potential energy surface are accessible 
(at a given temperature or energy) and indicates how a molecular system is prepared to react.
Secondly, the operator can directly observe the effect of structure manipulations on the molecular system as a whole.
This visual feedback relies on a continuously running real-time geometry optimization that drives the molecular system to nearby local minima.
Both types of feedback rely on quantum-mechanical forces calculated in real time by fast electronic structure methods.

For an optimal immersion into reactivity explorations, such real-time feedback needs to be reliable in different aspects.
From the operator's perspective, feedback must always be provided independently of the current state of the system, even if its exact calculation is not possible at all.
With respect to accuracy, feedback must give a qualitatively correct description of molecular behavior or, if such a description is unavailable, it must not 
lead the operator to inaccessible areas of configuration space.

Recently, we demonstrated the application of real-time reactivity studies described by the non-self-consistent density-functional tight-binding (DFTB) method.\cite{haag2014b}
DFTB\cite{porezag1995,seifert1996} is a non-iterative method featuring constant execution times and therefore delivers forces at a constant frequency.

Non-iterative methods such as DFTB are, however, too approximate to describe complex molecular systems reliably.
Hence, for most systems the application of self-consistent field (SCF) methods is required. 
However, the application of SCF methods introduces an additional layer of complexity due to their iterative nature.
With iterative methods, it cannot be guaranteed that the forces will be delivered at a constant frequency or that they will be delivered at all (if convergence cannot be achieved).

If, for a given structure, the execution time of the electronic structure optimization is incompatible with the real-time requirement, 
it will be necessary to freeze the molecular structure until the optimization converges as done by Luehr et al.\ in the field of 
steered \textit{ab initio} molecular dynamics.\cite{luehr2015a}
However, in the context of real-time reactivity explorations, immersion and hence interactivity would be jeopardized by freezing haptic and visual feedback. 

To maintain interactivity in real-time reactivity explorations based on iterative methods, the feedback cannot (and need not) reflect the exact quantum-mechanical forces directly because they might be provided only intermittently. 
In this work, we present a strategy that addresses this issue and allows performing interactive quantum-chemical reactivity studies with SCF methods.

\section{Timescales for Data Generation and Real-time Feedback}

In interactive chemical reactivity studies, information about the behavior of a molecular
system during operator-induced structure manipulations is conveyed to the operator by real-time rendering of data generated by electronic structure calculations. Rendering in this context entails the complete process of transforming raw data into data perceptible by human senses. 

To maintain interactivity, the rendering process must be performed in real time. 
Depending on the sense addressed by the rendering, different real-time requirements must be fulfilled. 
Addressing the visual and haptic sensory systems impose different requirements for the update rate that is to be maintained. 
Achieving certain minimum update rates translates into maximum time intervals that are allowed for the data generation and rendering process. 

The time interval available for the calculation of a result is therefore defined by the senses addressed by the feedback mechanism. 
For a setting where feedback consists of rendering molecular forces and visualizing structural changes upon manipulations by the operator, Table~\ref{tab:senses} gives an overview of the imposed time limits and the types of calculation to be performed in the available time. 

\setlength{\tabcolsep}{6pt}
\begin{table}[h]
	\centering
		\begin{tabular}{l c c}
			\hline
			\hline
			                      & visual  & haptic  \\
			\hline
			Frequency                  & 60 Hz & 1 kHz \\
			Time interval           & 16 ms & 1 ms \\
			Calculation	               & structure update & force calculation \\
			\hline
			\hline
		\end{tabular}
	\caption{Senses addressed in the haptic quantum chemistry setting and their implications for quantum-chemical calculations.}
	\label{tab:senses}
\end{table}
\setlength{\tabcolsep}{4pt}

Basically all reliable electronic structure methods require an iterative optimization of the electronic wave function or density (e.g., an SCF
optimization of the orbitals). 
The iterative nature of these methods, however, leads to unpredictable execution times. 
Assuming that all parameters of a calculation are fixed and only the molecular structure varies from calculation to calculation, as is the case in explorations of chemical reactivity, it cannot be known \textit{a priori} whether and when a calculation will converge. 
This also holds for fast methods that usually converge in few iterations but might not converge at all for far-from-equilibrium structures generated by the operator.

\begin{figure}[htb]
\begin{center}
 \includegraphics[width=0.48\textwidth]{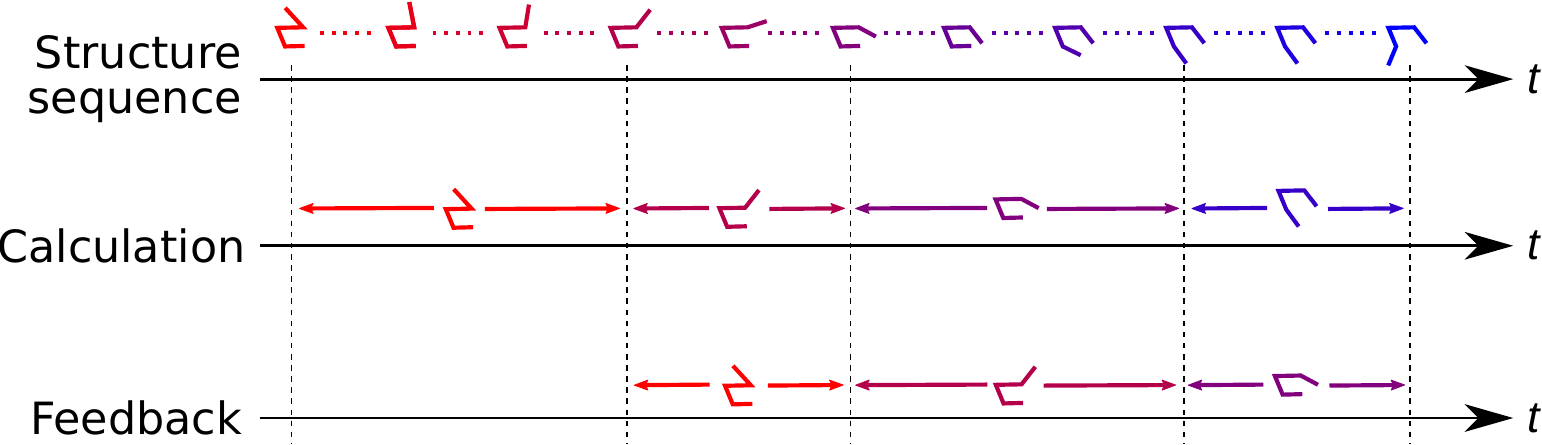}
\caption{
	Changing molecular structure (depicted by colored shapes) of a real-time exploration.
	Top: Real-time evolution of the molecular structure as displayed to the operator. The molecular structure is updated with 60~Hz.
	Middle: Electronic structure calculations are performed only for some of the visited structures.
	Bottom: The feedback relies on molecular structures for which the calculations finished.
	The vertical dashed lines represent instants of time when an electronic structure calculation finished and another one starts.
\label{fig:delay}}
\end{center}
\end{figure}

The finite execution time of electronic structure calculations introduces, at a given instant of time, shifts in consistency between the molecular structure displayed to the operator, the structure used for the data generation, and the structure to which the feedback corresponds. 
In practice, the electronic structure can be optimized only for a fraction of the molecular structures visited 
as soon as the electronic structure calculation cannot be performed at 60~Hz.
The feedback is always given with a certain delay because of the finite execution time of the electronic structure optimization and because of 
the time elapsed between the completion of the last electronic structure optimization and the instant of time at which the feedback is needed. 
This is represented schematically in Fig.~\ref{fig:delay}.

The delay will not be perceptible if the electronic structure calculations are very fast, but it can lead to artifacts in the operator's perception as soon as their execution time increases. 
In Fig.~\ref{fig:delay}, a new electronic structure calculation starts only when the previous one is finished. 
Starting parallel electronic structure calculations for consecutive molecular structures can reduce the delay by updating the data on which the feedback relies more frequently, but the delay will never vanish. 

The inherently unpredictable execution times and the non-vanishing delays even in the most well-behaved situations prohibit the direct 
application of iterative electronic structure calculations as a data source for real-time feedback as required in interactive applications.

\section{Mediator strategy}

In a setting that allows an operator to interact with virtual objects, the response of the system to the operator's actions is presented by a feedback
component. This feedback component is usually fed with data by some data generation component---usually some sort of calculation. Providing 
real-time feedback imposes strict deadlines for the provision of the data. Hence, the data flow from generator component to feedback component can only 
be direct if the data generation is performed in a predictable and constant time interval. 
If, however, the update rates of data generation and feedback component differ, filtering techniques must be utilized to render the incoming data stream.

\begin{figure}[h]
	\centering
		\includegraphics[width=\columnwidth]{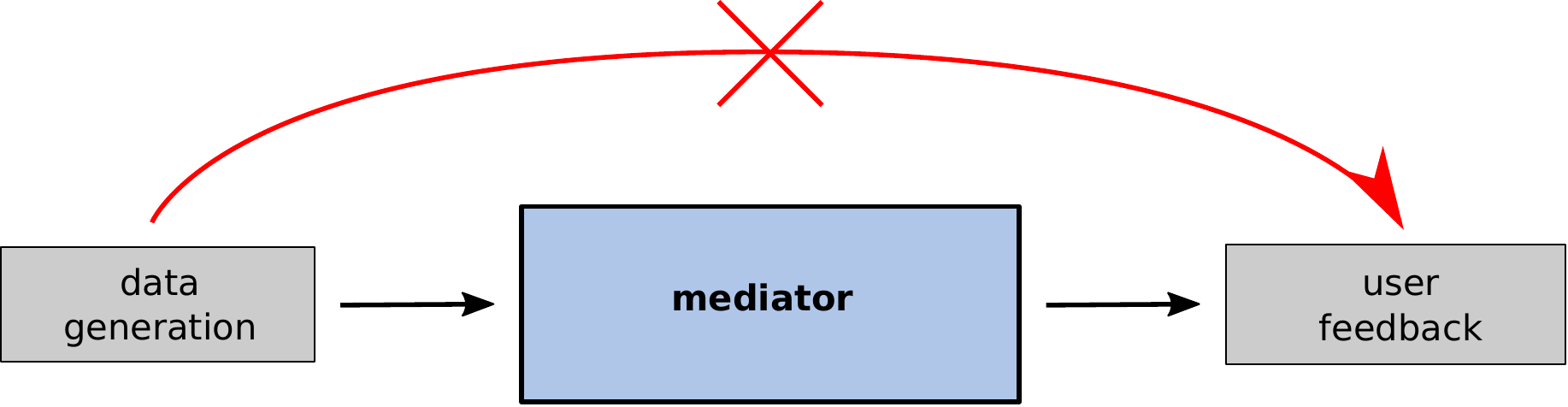}
	\caption{The three main components necessary for interactive systems with real-time feedback and unpredictable data availability.}
	\label{fig:Mediator}
\end{figure}

If the data generator cannot guarantee the provision of new data in fixed time intervals, an additional component is necessary to mediate
between the unpredictable data output stream and the strict real-time requirements imposed by the operator feedback (see Fig.~\ref{fig:Mediator}). The 
mediator consumes the data stream coming from the generating calculation and, based on it, provides real-time data for the operator feedback. 
The mediator must fulfill three important requirements:
\begin{itemize}
	\item It must guarantee to deliver data for the feedback in fixed and constant time intervals (whose value is set by the human sense addressed).
	\item It must provide reliable feedback in any situation.
	\item The data output must be based on the most recent data provided by the data generator. 
\end{itemize}

The mediator in turn must rely on the ability of the data generator to provide a time ordering of the data stream. 
This means that the raw data must be assigned to a certain stage in the exploration so that the mediator is able to provide an adequate response. 
The time ordering will automatically be achieved if the data generator performs the calculations consecutively. 
In the case of parallel calculations, the data generator must provide additional information on what results belong to which instant of time. 

A trivial implementation of the mediator would be to provide the last data received to the operator feedback component until it receives 
new data from the generator. However, as the most recent data will no longer correspond to the situation the operator is experiencing (mismatch of property data for a user-modified new structure), interactivity will be severely hampered or will be lost entirely. 
Therefore, this trivial implementation does not fulfill the second requirement for a mediator.

\subsection{The Surrogate Function}

When the feedback is a function of the current configuration of the system, a mediator will be required
if the time needed for the evaluation of the function is large or unpredictable.
In such cases, a mediator fulfilling the above requirements can be implemented in terms of a surrogate function that can be evaluated rapidly.
We will refer to the unknown but correct function as 'reference' or 'exact' function. 

The reference function is still sampled as often as possible and delivers data allowing for updating the shape of the surrogate function.
The surrogate function can be evaluated at the high frequencies required by the feedback component.

\begin{figure}[h]
	\centering
		\includegraphics[width=\columnwidth]{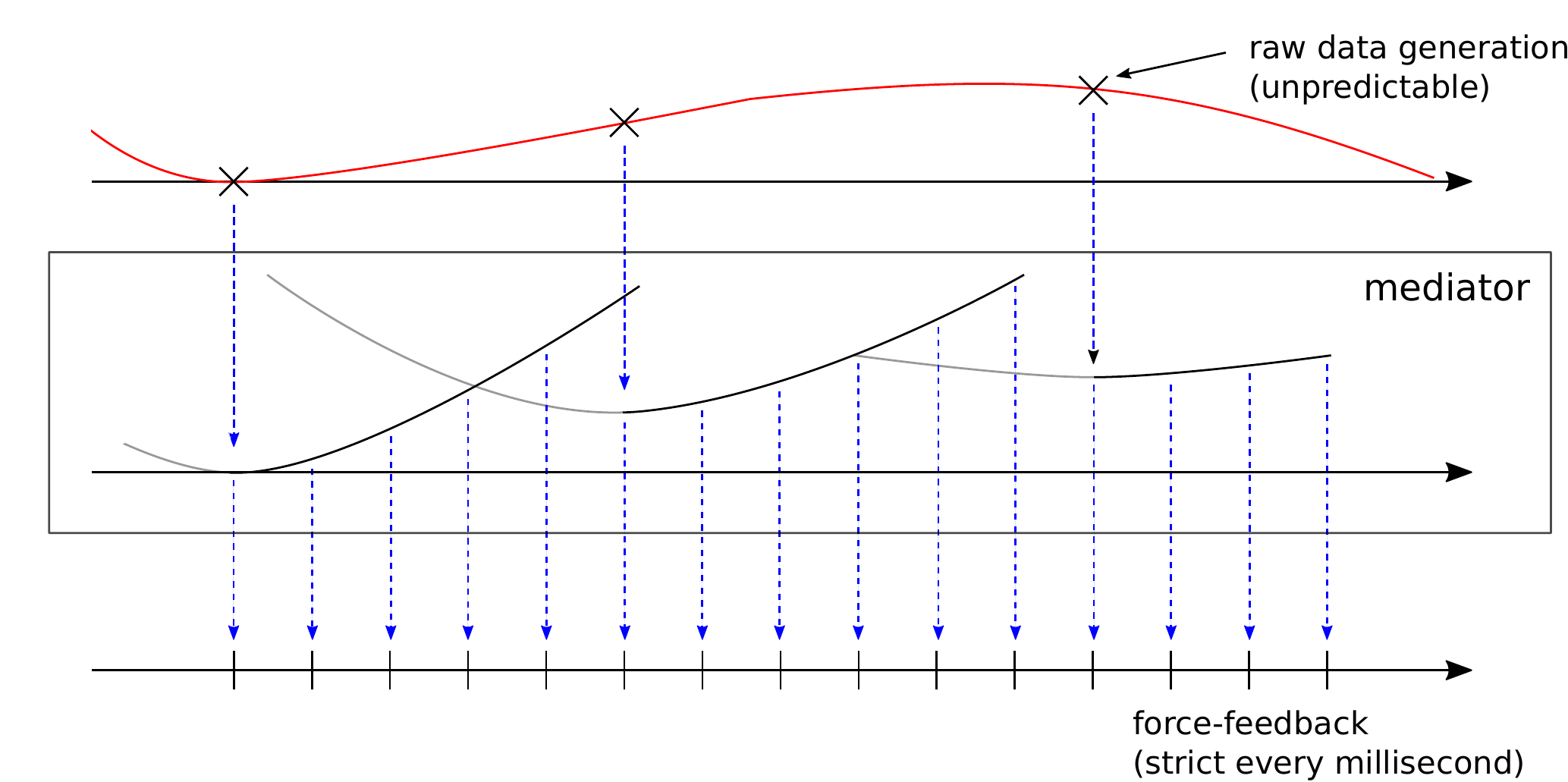}
	\caption{The reference potential (red) is sampled (blue arrows) to generate the surrogate potentials (black), which are in turn sampled to provide 
	the real-time feedback data. Unused parts of the surrogate potentials due to the arrival of new reference data are indicated in gray. }
	\label{fig:SurrogatePotential}
\end{figure}

If the reference function is a potential or energy and the real-time feedback relies on forces, we will call the surrogate function
a 'surrogate potential'. 
The surrogate potential is chosen to approximate the local shape of the reference potential.
The force feedback is evaluated from the surrogate potential, as illustrated in Fig.~\ref{fig:SurrogatePotential}.

For the surrogate potential as the central component of the mediator, 
it is not important how new reference calculations are started as long as the results are strictly ordered in time. 
The mediator only needs to know which data is most recent. 
The calculations can therefore be performed consecutively (a new calculation starts only when the previous one finished) or in parallel (a new calculation can start before the previous one finished). 

The requirement on the mediator to always yield reliable data that prevents the operator to explore unreasonable areas of configuration space translates into the requirement for the surrogate potentials to be always bounded from below. 
Such surrogate potentials ensure conservative forces that will drive the operator toward a guess for a local minimum of the reference potential, which usually also corresponds to the part of the reference potential well reproduced by the surrogate potential.

It must be emphasized that the surrogate potential is not an interpolation scheme as it uses only local information from the most recent calculation to 
predict the reference potential until new data arrives. An interpolation of the accumulated exploration history would in most cases not improve the 
prediction as most of the time the operator's manipulations will explore new areas of configuration space that cannot be foreseen.

\subsection{Obtaining the Surrogate Potential for Feedback from Electronic Structure Calculations}

For real-time quantum chemistry, the reference potential can, for instance, 
be approximated by surrogate potentials of quadratic form, which allow for an efficient calculation of the forces needed for the structure relaxation and for the haptic force rendering.

\begin{figure}[!htb]
\begin{center}
 \includegraphics[width=0.48\textwidth]{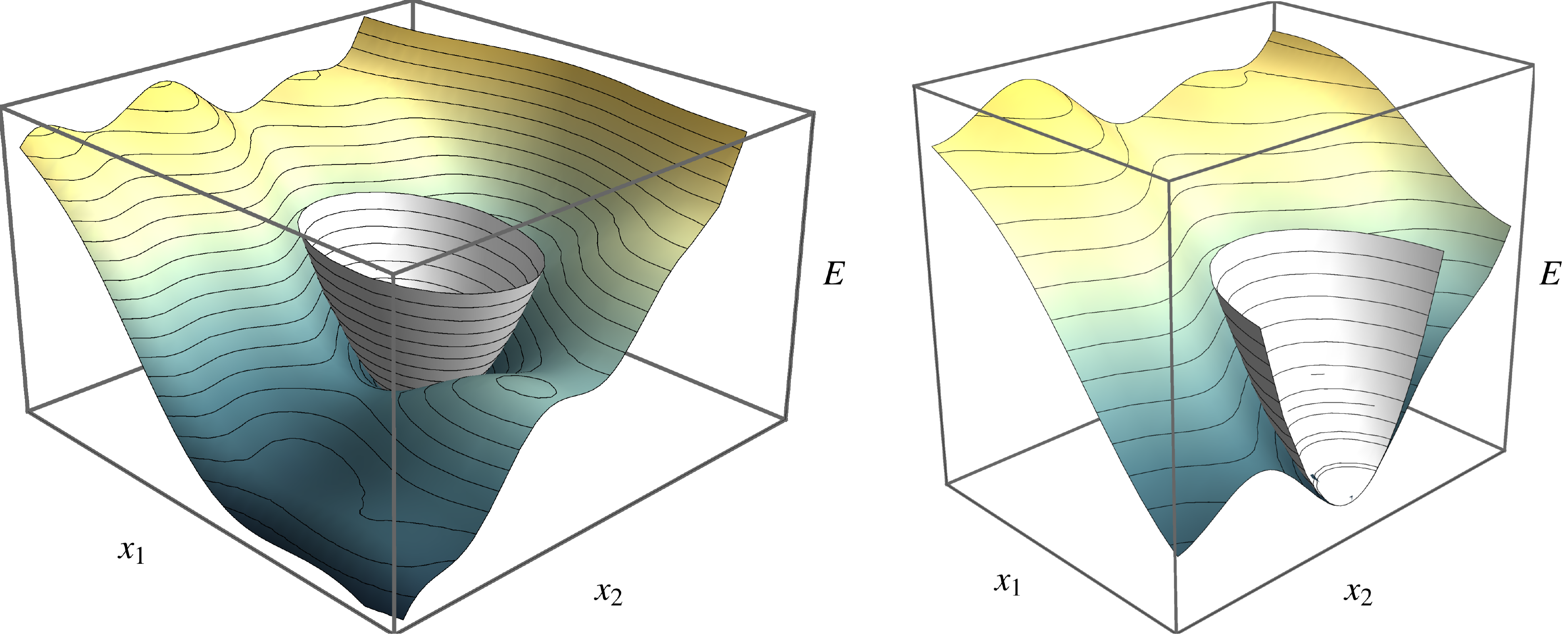}
\caption{
	Left: Arbitrary potential energy surface in two dimensions and the surrogate potential approximating it. 
	Right: Close-up representation.
\label{fig:surrogate}}
\end{center}
\end{figure}

Quadratic potentials are characterized by the general expression

\begin{align}
	V_\mathrm{sur}(\boldsymbol{x}) = V_0 &+ \boldsymbol{a}^T (\boldsymbol{x}-\boldsymbol{x}_0) \nonumber\\
	&+ \frac{1}{2} (\boldsymbol{x}-\boldsymbol{x}_0)^T \boldsymbol{B} (\boldsymbol{x}-\boldsymbol{x}_0)
\label{eq:paraboloid}
\end{align}

where $\boldsymbol{x} = \{ x_1, \dots , x_m \}$ is a coordinate vector for the atomic (i.e., nuclear) 
positions, $V_0$ a constant potential shift, $\boldsymbol{a}$ an $m$-dimensional vector and $\boldsymbol{B}$ an $m \times m$ matrix.
$\boldsymbol{x}_0$ is a reference position and, in the context of real-time quantum chemistry, will correspond to the configuration at which the reference potential was sampled last.
Fig.~\ref{fig:surrogate} illustrates the quadratic approximation of a given potential in two-dimensional configuration space ($m=2$).

Quadratic potentials allow for a straightforward evaluation of the forces,
\begin{align}
	\boldsymbol{F}_\mathrm{sur} (\boldsymbol{x}) = - \boldsymbol{\nabla} V_\mathrm{sur}(\boldsymbol{x}) 
	= -\boldsymbol{a} - \boldsymbol{B}(\boldsymbol{x}-\boldsymbol{x}_0),
\label{eq:forces}
\end{align}
provided that $\boldsymbol{B}$ can be efficiently calculated.
$\boldsymbol{F}_\mathrm{sur}$ is a vector composed of atomic forces and represents the basis for real-time force feedback.

To reproduce the reference potential around the atomic positions as accurately as possible, the surrogate potential is chosen to reproduce the local curvature of the potential around the atomic nuclei.
Given a molecular structure for which the electronic wave function was optimized, the coefficients of the surrogate potential around $\boldsymbol{x}_0$ are derived from the electronic energy $E$ and its derivatives as:
\begin{align}
\label{eq:coef1}
	V_0 &= E(\boldsymbol{x}_0), \\
\label{eq:coef2}
	a_i &= \frac{\partial E(\boldsymbol{x})}{\partial x_{i}} \bigg|_{\boldsymbol{x} = \boldsymbol{x}_0}, \\
\label{eq:coef3}
	B_{ij} &= \frac{\partial^2 E(\boldsymbol{x})}{\partial x_{i} \partial x_{j}} \bigg|_{\boldsymbol{x} = \boldsymbol{x}_0}.
\end{align}

Applying Eqs.\ (\ref{eq:coef1})--(\ref{eq:coef3}) will occasionally generate quadratic potentials with maxima or saddle points, which will not be bounded from below. 
Such potentials are characterized by $\boldsymbol{B}$ not being positive definite, i.e., $\boldsymbol{B}$ then possesses one or several negative eigenvalues. 
To change them into potentials with a global minimum, i.e., into bounded potentials, can be achieved by replacing all negative 
eigenvalues of $\boldsymbol{B}$ by their absolute values:
\begin{enumerate}
	\item Diagonalize $\boldsymbol{B}$ to yield $\boldsymbol{B}_\mathrm{diag}=\boldsymbol{U}^{-1} \boldsymbol{B} \boldsymbol{U} = \{B_{\mathrm{diag},ij}\}$, \\ 
		$B_{\mathrm{diag},ij} = b_{ij} = \begin{cases}
			0 & i \neq j \\
			b_i & i=j\\
		\end{cases}$
	\item In $\boldsymbol{B}_\mathrm{diag}$, replace all negative eigenvalues by their absolute value, $b_i'=\mathrm{abs}(b_i)$, to generate the matrix $\boldsymbol{B}_\mathrm{diag}'$. 
	\item Back-transform to a new matrix $\boldsymbol{B}' = \boldsymbol{U}\boldsymbol{B}_\mathrm{diag}'\boldsymbol{U}^{-1}$ with positive eigenvalues $b_i'$.
\end{enumerate}

\begin{figure}[!htb]
\begin{center}
 \includegraphics[width=0.48\textwidth]{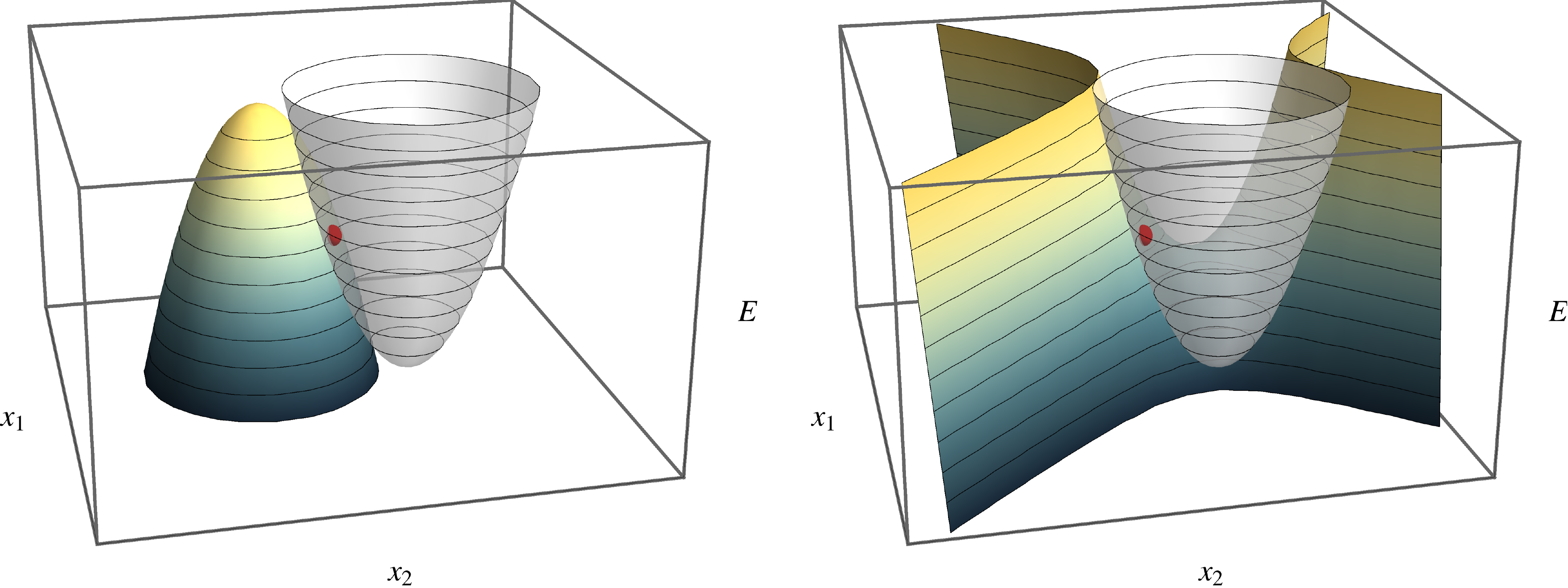}
\caption{
	The reference quadratic potential, possessing a maximum (left) or a saddle point (right), is changed into the potential
        printed in light gray, which strictly posseses a minimum.
	The reference point around which the conversion is performed is represented by a red dot.
\label{fig:transformation}}
\end{center}
\end{figure}

The gradient at the reference position $\boldsymbol{x}_0$ is preserved and the original stationary point is changed into a minimum. 
The effect of converting an unbounded quadratic potential into a bounded quadratic potential is illustrated in Fig.~\ref{fig:transformation} for two dimensions.

For a molecular system consisting of $N$ atoms, it is possible to apply quadratic surrogate potentials in two ways. 
First, one can approximate the reference potential by a single surrogate potential in $m=3N$ dimensions for all degrees of freedom of the system. 
In this case, $\boldsymbol{B}$ is the Hessian matrix. 
Second, the reference potential can be approximated by a combination of $N$ quadratic potentials in $m=3$ dimensions, each depending on one atomic position. 
The second option corresponds to approximating the Hessian matrix by its block-diagonal and will therefore be less accurate. 
For real-time quantum chemistry, this approximation is tolerable because structure modifications are mainly local.
The second option is computationally more efficient as it involves the calculation of $6N$ second derivatives instead of $(3N)(3N+1)/2$ and does not require the diagonalization of a $3N \times 3N$ matrix, but of $N$ $3 \times 3$ matrices.
Both approaches can be applied in real-time quantum chemistry. 

For the haptic force rendering, a quadratic potential of dimension $m=3$ allows for a fast evaluation depending on the coordinates of the manipulated atom only. 
Therefore, a force feedback of 1 kHz can be guaranteed even if the electronic structure calculations finish at a lower rate. 
This results in an excellent responsiveness upon manipulation. 

\begin{figure}[htb]
\begin{center}
 \includegraphics[width=0.32\textwidth]{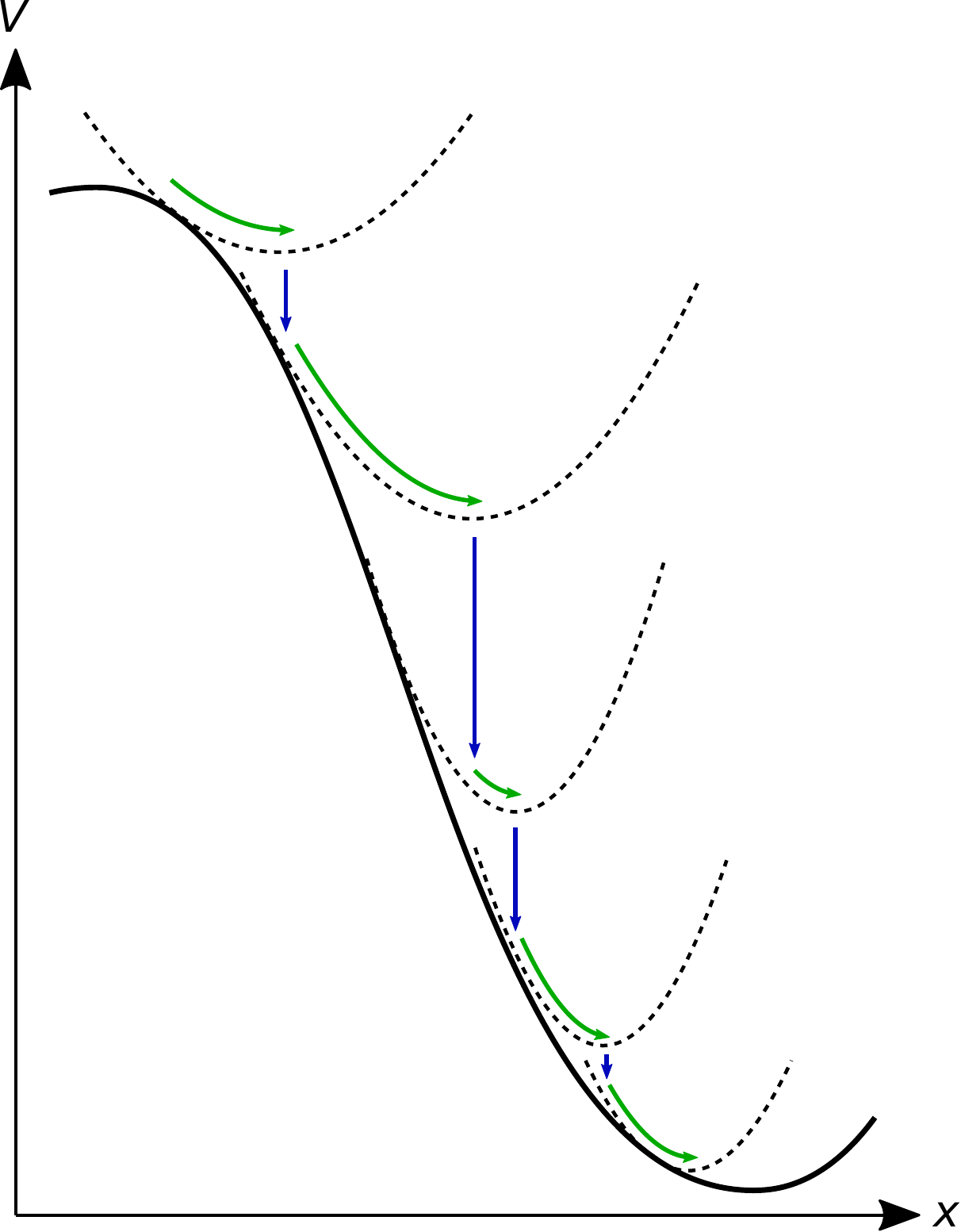}
\caption{
	Trajectory of an atomic nucleus following a steepest-descent relaxation in real time. 
	The black line represents the reference potential for the atom, which is in parts unknown in the course of an interactive exploration.
	The structural evolution according to the steepest-descent algorithm is based on the surrogate potentials shown by dashed parabolas.
	The motion of the atom (green arrows) occurs toward the minimum of the current surrogate potential. 
	When the surrogate potential is updated (blue vertical arrows), the motion can proceed again.
\label{fig:structuralRelaxation}}
\end{center}
\end{figure}

The surrogate potentials also provide all necessary forces for the structure response (e.g., steepest-descent structure relaxation). 
If only the first derivatives of the potential energy were used for the relaxation, atoms could easily move so far during the time 
required for the electronic structure optimization that explicitly calculated forces and molecular structures hardly correspond. 
Upon structure changes, the surrogate potentials limit the motion of atoms until the next electronic structure calculation converges 
because the quadratic potentials can be guaranteed to feature one single minimum.
A schematic representation of the effect of the application of surrogate potentials in the context of structure relaxation is shown in Fig.~\ref{fig:structuralRelaxation}.

\begin{figure}[!htb]
\begin{center}
\includegraphics[width=0.48\textwidth]{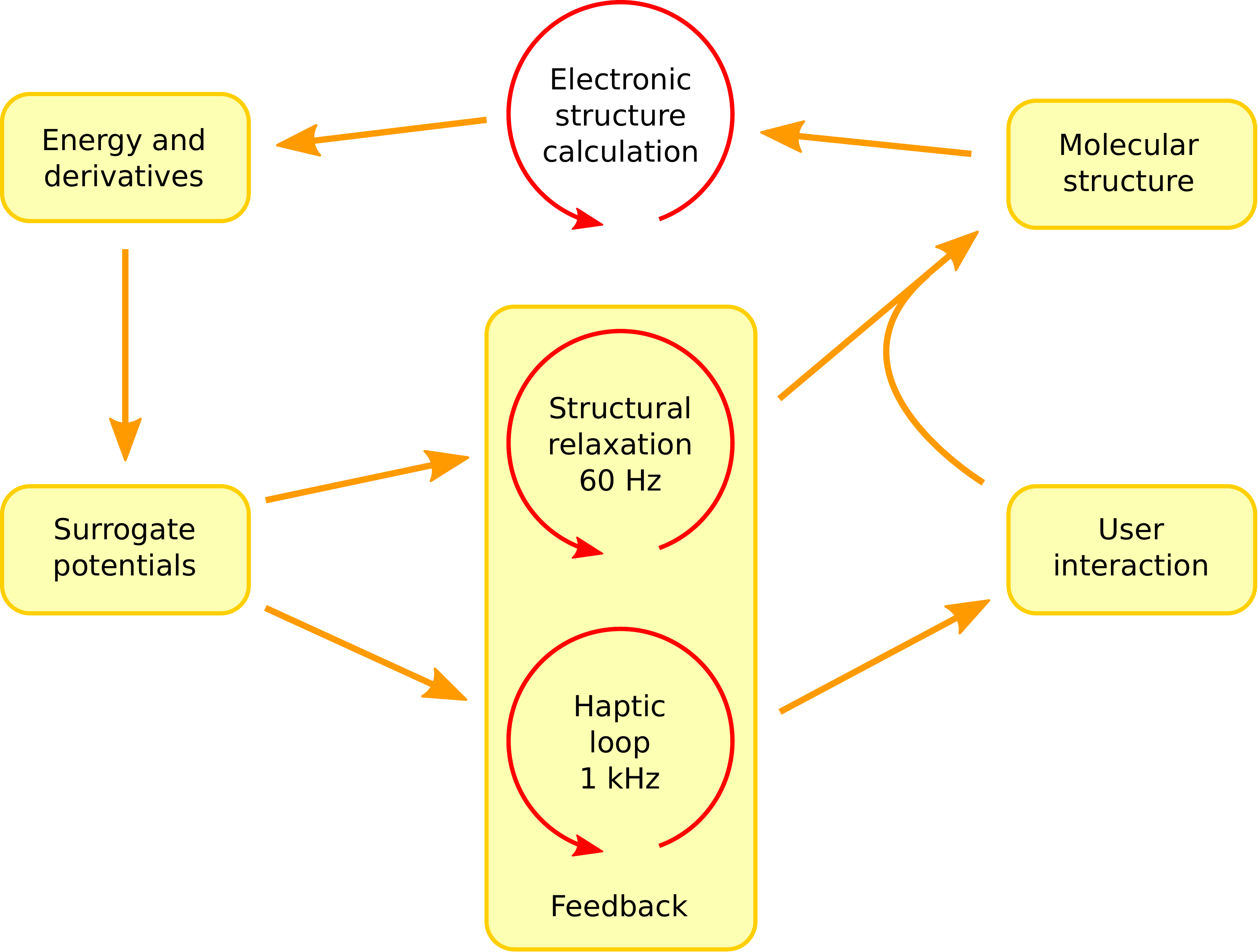}
\caption{
	Schematic flow of force generation in real-time quantum chemistry. 
	The loops are represented by curved red arrows and the information flow by orange arrows.
\label{fig:DataFlow}}
\end{center}
\end{figure}

The implementation of surrogate potentials in real-time quantum chemistry is represented schematically in Fig.~\ref{fig:DataFlow}. 
Three independent loops (Fig.\ \ref{fig:DataFlow}, middle) run continuously. In the first one, electronic structure calculations (of unpredictable execution time) are performed successively. 
This loop fetches the current molecular structure before each electronic structure calculation starts and generates the surrogate potentials from the energy and its derivatives when it finishes. 
The second loop runs at 60~Hz and is responsible for the structure relaxation of the system based on the steepest-descent algorithm. 
The third loop runs at 1~kHz. It tracks the structure manipulations by the operator and renders a force feedback through the haptic device.

\section{Numerical Results}

We extended our real-time quantum chemistry framework\cite{haag2014b} within the {\sc SAMSON}\cite{bosson2012b} molecular editor to reactivity studies 
based on iterative electronic structure methods. 
As SCF methods, we implemented the semi-empirical Parametrized Method 6 (PM6)\cite{stewart2007} and self-consistent DFTB variants \cite{elstner1998,gaus2011}. 
This implementation allowed us to assess the application of mediators in the form of surrogate potentials.
In the following, we focus on PM6 as the data source.
The PM6 reference potential is approximated by a combination of three-dimensional surrogate potentials, one for each atom of the molecular system under consideration.
The surrogate potentials are generated from the PM6 energy, first derivatives and frozen-density second derivatives.
Frozen-density second derivatives are obtained by assuming a constant density matrix in the expression for the analytic second derivatives.
The generation of potentials from exact analytical second derivatives has not been considered because their calculation involves the solution of the coupled perturbed Hartree--Fock equations and is computationally significantly more expensive than the calculation of the gradients, even for semi-empirical methods.\cite{frisch2009a}

\begin{figure}[!h]
\begin{center}
\includegraphics[width=0.4\textwidth]{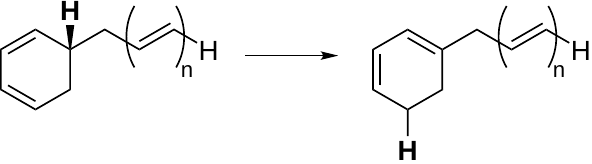}
\caption{
	[1,5] hydride shift reaction.
	The side chain length depends on the value of the integer $n$.
\label{fig:reaction}}
\end{center}
\end{figure}

To assess the reliability of surrogate potentials for real-time feedback, the real-time exploration of a [1,5] hydride shift reaction with variable side chain length, shown in Fig.~\ref{fig:reaction}, was studied with and without surrogate potentials. 

The variable size of the side chain allows us to study different molecular sizes with different average execution times of the electronic structure optimization.
For the data production, the hydrogen atom displayed in bold in Fig.~\ref{fig:reaction} was moved in a straight line at constant speed.

\subsection{Haptic Force}
\begin{figure*}[p]
\begin{center}
\includegraphics[width=0.93\textwidth]{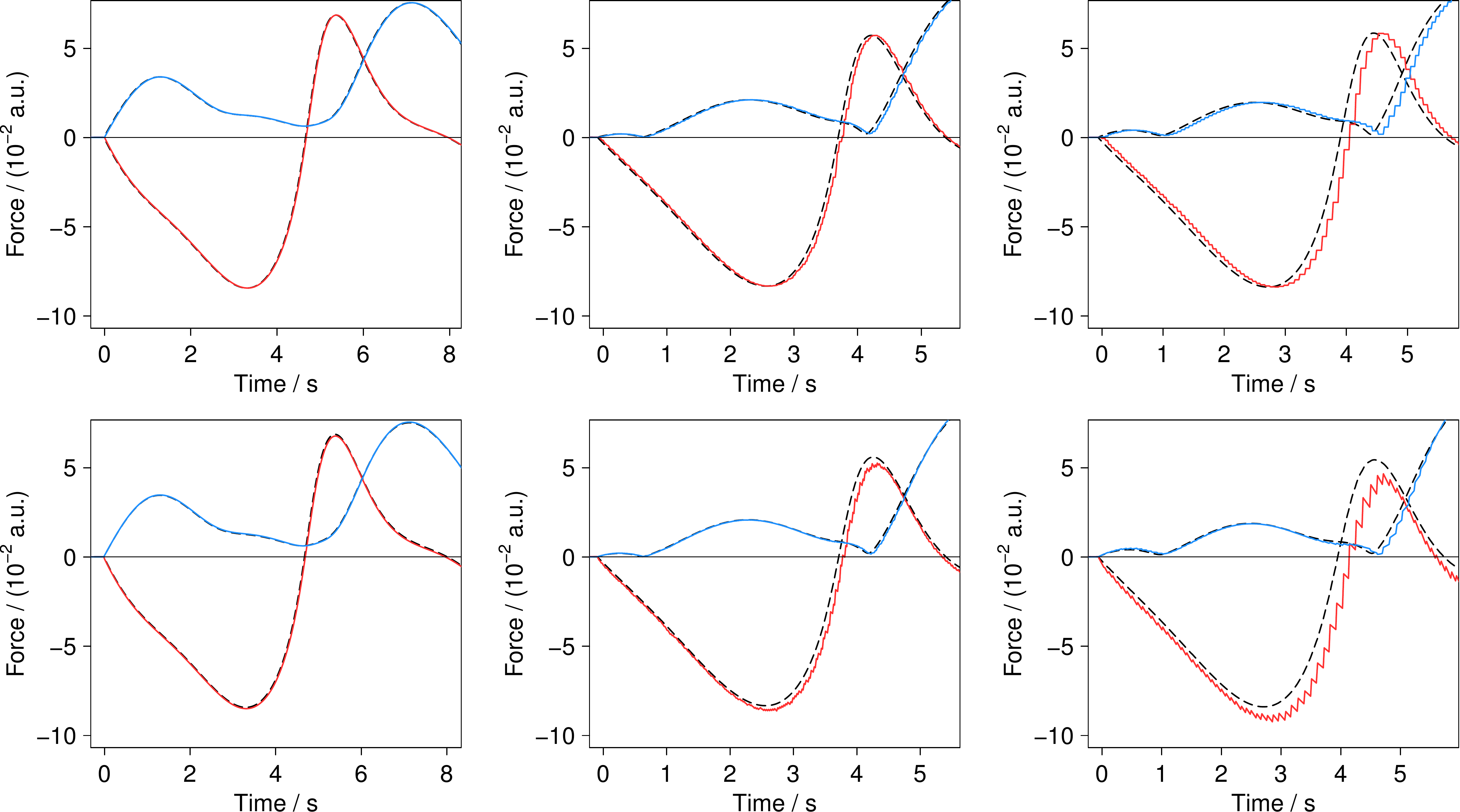}
\caption{
	Parallel (red) and perpendicular (blue) forces calculated in real time for the hydrogen atom involved in the hydride shift reaction of Fig.~\ref{fig:reaction}. 
	The exact PM6 forces are depicted by black dashed lines.
	The motion of the hydrogen atom was initiated at time $t=0\,\mathrm{s}$.
	Top row: No surrogate potentials are applied.
	Bottom row: Surrogate potentials are applied.
	Left column: $n=0$;
	Middle column: $n=8$;
	Right column: $n=16$, with $n$ being the number of double bonds in the side chain for the model reaction in Fig.~\ref{fig:reaction}.
\label{fig:plots}}
\end{center}
\end{figure*}

To evaluate the effect of surrogate potentials on the haptic force, the forces acting on the hydrogen atom during the reaction were recorded. 
The structures visited during the reaction were stored for the calculation of the exact PM6 forces after the exploration. 
Exact PM6 forces and real-time forces were then decomposed into parallel and perpendicular components with respect to the direction of motion of the hydrogen atom. 
The results with and without surrogate potentials for the model reaction with $n=0,8,16$ are presented in Fig.~\ref{fig:plots}.
For the structure relaxation, the steepest-descent algorithm with $\gamma = 0.1\,(\mathrm{bohr})^2/\mathrm{Hartree}$ was applied 
($\gamma$ is defined in Eq.\ (\ref{gammaeq}) below).

\begin{figure*}[p]
\begin{center}
\includegraphics[width=0.93\textwidth]{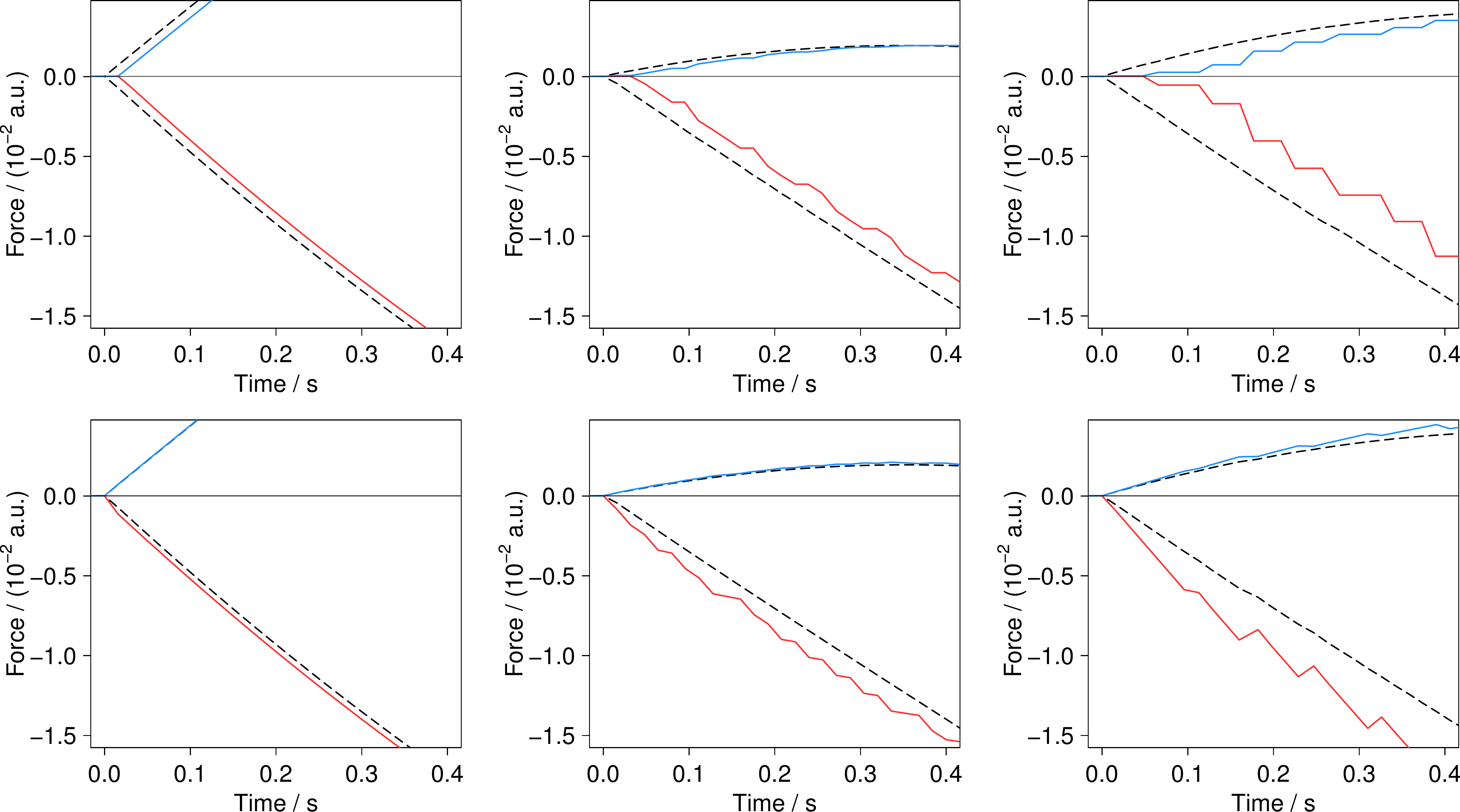}
\caption{
	Enlargement of the graphs in Fig.~\ref{fig:plots} displaying the rendered forces shortly after the motion of the hydrogen is initiated.
        Notation as in Fig.~\ref{fig:plots}. 
\label{fig:plots_onset}}
\end{center}
\end{figure*}

For small molecular systems with short calculation times (left column in Fig.~\ref{fig:plots}), the exact PM6 forces are reproduced almost exactly at all 
times.
The differences resulting from the introduction of surrogate potentials become more pronounced for longer execution times of the electronic structure optimization (middle and right columns in Fig.~\ref{fig:plots}).
Without surrogate potentials, the applied forces correspond to earlier structures, which results in a temporal shift that becomes larger when the SCF procedure needs more iterations. 
With a quadratic surrogate potential, the surrogate forces are piecewise linear since the motion of the hydrogen atom is constant.
Importantly, the surrogate potential allows for an immediate force feedback when the motion of the hydrogen atom is initiated, as shown in Fig.~\ref{fig:plots_onset}.
This is a consequence of the increased responsiveness of force feedback upon sudden changes of exploration direction, made possible by surrogate potentials.

As mentioned above, the purpose of the surrogate potential is to deliver reliable forces in unknown areas of configuration space and \textit{not} to exactly match the reference potential.
Although the forces appear to be, on average, less accurate when the surrogate potential is applied, Fig.~\ref{fig:plots} shows that this requirement is fulfilled even when the surrogate potential cannot be updated at a high frequency.
Furthermore, the provided forces are conservative as they are stronger than the reference when they are opposing the manipulation by the operator 
and smaller than the reference when they are in the same direction. 

\subsection{Structural relaxation}
Our framework allows for structural relaxation by applying the steepest-descent algorithm with a frequency of 60~Hz.
Note that other algorithms can be applied as well and lead to similar conclusions.
In a steepest-descent step, the position $\boldsymbol{x}_i(t_n)$ of each atom $i$ evolves along the atomic force $\boldsymbol{F}_i(t_n)$:
\begin{align}
\label{gammaeq}
	\boldsymbol{x}_i(t_{n+1}) = \boldsymbol{x}_i(t_n) + \gamma \boldsymbol{F}_i(t_n)
\end{align}
The application of surrogate potentials has a direct effect on the maximal value of $\gamma$ that can be chosen and therefore on the maximal relaxation speed.
Table~\ref{tab:stepsize} shows, for the model reaction of Fig.~\ref{fig:reaction} with $n=0,8,16,24$, the maximal value of $\gamma$ that can be employed without leading to artifacts such as artificial oscillations of atomic positions or tearing apart the molecular system. 

\begin{table}[!h]
	\centering
\begin{tabular}{l c c c c}
	\hline
	\hline
	 & $n=0$ & $n=8$ & $n=16$ & $n=24$ \\
	\hline
	w/o $V_\mathrm{sur}$  & $1.35$ & $0.45$ & $0.15$ & $0.08$ \\
	w $V_\mathrm{sur}$ & $1.35$ & $1.45$ & $1.50$ & $1.50$ \\
	\hline
	\hline
\end{tabular}
\caption{Maximal value of $\gamma$ (in $(\mathrm{bohr})^2/\mathrm{Hartree}$) applicable for the model reaction of Fig.~\ref{fig:reaction} with and without surrogate potentials.
	The structural relaxation was performed at $60$~Hz and $n$ is the number of double bonds in the side chain for the model reaction of Fig.~\ref{fig:reaction}.}
	\label{tab:stepsize}
\end{table} 

Without surrogate potentials, the maximal value of $\gamma$ needs to be adapted to the system size. 
The larger the molecular structures, the longer the execution time of the electronic structure optimization is 
and, at constant structure-relaxation frequency, the more steps are to be performed based on the same forces. 
Applying surrogate potentials decouples $\gamma$ and the size of the system, and therefore allows for larger values of $\gamma$.
In this case, the maximal values for $\gamma$ rather depend on the local shape of the potential, i.e., of the second derivative of the surrogate potential, and on the frequency at which the structure relaxation is performed. 
By default, our framework will employ $\gamma = 0.5\,(\mathrm{bohr})^2/\mathrm{Hartree}$ independent of the molecular system size if 
surrogate potentials are applied.

\section{Conclusion}

A solution is presented to mediate between the inherently unpredictable execution times of iterative electronic structure calculations and the 
time requirements of real-time feedback. 
In the context of real-time quantum chemistry, such feedback relies on quantum-mechanical forces and must be generated at
high frequencies that cannot be guaranteed when iterative electronic structure methods are applied. 

The strategy followed in this work is to calculate the atomic forces from surrogate potentials that can easily be evaluated for molecular structures 
for which the electronic structure is still unknown.
For consecutive molecular structures visited during real-time reactivity explorations, one or several surrogate potentials are generated from the electronic energy and its first and second derivatives to approximate the reference potential energy surface. 
Surrogate potentials such as bounded quadratic potentials allow for a fast evaluation of forces compatible with real-time requirements.

To calculate forces at such high frequencies offers immediate benefits for real-time feedback in the form of on-the-fly structure relaxation and haptic force feedback.

In the context of structure relaxation, surrogate potentials avoid the need for special measures such as freezing the system when electronic structure calculations do not converge or are very time-consuming --- the molecular structure will evolve toward the minimum of the surrogate potential and remain there until the surrogate potential is updated.
Consequently, the structural evolution can be performed faster (translated into larger steps in the steepest-descent algorithm) since the surrogate potentials guarantee to lead to a minimum.
In particular, the step size of the structure relaxation algorithm does not need to be adjusted for different system sizes or calculation times.

For haptic feedback, the ability to update forces at 1~kHz improves substantially on the responsiveness of force feedback upon structure manipulations by the operator. 
Furthermore, in the case of non-converging calculations, the surrogate potential will drive the operator back to its minimum and hinder large manipulations from his part when insufficient information about the reference forces is available.

The presented procedure is independent of the system size and is also advantageous when the execution time of the electronic structure calculation is constant and predictable.
Despite the additional computational time needed for the calculation of (approximate)
second derivatives, the application of surrogate potentials proves necessary and beneficial for interactive reactivity studies based on iterative methods.

\section*{Acknowledgments}
This work was generously supported by ETH Research Grant ETH-20 15-1 and ETH Pioneer Fellowship Grant PIO-11-14-2.


\end{document}